\documentclass[prb,aps,twocolumn,nofootinbib,superscriptaddress]{revtex4-1}
\usepackage{amsmath}
\usepackage{amsfonts}
\usepackage{amssymb}

\newcommand{\vect}[1]{\mathbf{#1}}
\usepackage{amsmath}
\usepackage{amsfonts}
\usepackage{graphicx}
\usepackage{verbatim}
\newcommand{\Z}{\mathbb{Z}}

\begin{document}
\title{Superconductivity with intrinsic topological order\\
induced by pure Coulomb interaction and time-reversal symmetry breaking}

\author{Evelyn Tang}
\author{Xiao-Gang Wen}
\affiliation{Department of Physics, Massachusetts Institute of Technology, Cambridge, MA 02139}
\affiliation{Perimeter Institute for Theoretical Physics, Waterloo, Ontario, N2L 2Y5 Canada}

\date{May, 2013}

\begin{abstract}
Recently, in certain flat band lattice systems at commensurate fillings, fractional quantum Hall states have been found -- which have anyonic excitations. We study such systems away from commensuration, i.e. the ground state of an anyon gas in such a system. The presence of the underlying lattice allows access to an entirely new regime where the anyon kinetic energy can be larger than their interaction energy. Within the flux-attachment approach, using mean-field then adding fluctuations, we find several possible superfluid states. Two have intrinsic topological order, i.e.  fractionalized quasiparticles with a fusion structure of $(\Z_2)^4$ and
$(\Z_8)^2$ respectively, and a third has no fractionalized excitations similar to a BCS-type state. This represents a mechanism for superconductivity driven purely by strong repulsion
and complex hopping of electrons.

\end{abstract}

\maketitle
\section{Introduction}
Recently, there have been proposals for the fractional
quantum Hall (FQH) effect to be realized in a lattice system without magnetic
field and at high temperatures, for instance in a flat band with spin-orbit coupling and spin
polarization
\cite{TMW1106,SGK1103,NSC1104}. At commensurate filling fractions of this flat
band, i.e. the electron
number per unit cell is a simple rational number e.g. 1/3, it has been shown numerically that the ground-state in such
systems is a FQH state \cite{SGS1189,RB1114}. A natural question is what
happens at incommensurate fillings, when the electron density is doped
away from a rational fraction and a gas of anyon excitations created.

The presence of the underlying lattice system allows us to access an entirely
new regime where the anyon excitations may have kinetic energy larger than
their interaction energy. This is in contrast to the FQH state in semiconductor
systems, where electrons have zero bandwidth and anyons have a magnetic length
scale several orders of magnitude larger than in lattice systems. Consequently,
the anyon
is expected to have very little dispersion, favoring localization or Wigner
crystal formation.

\begin{figure}[tb]
\begin{center}
\includegraphics[width=\linewidth]{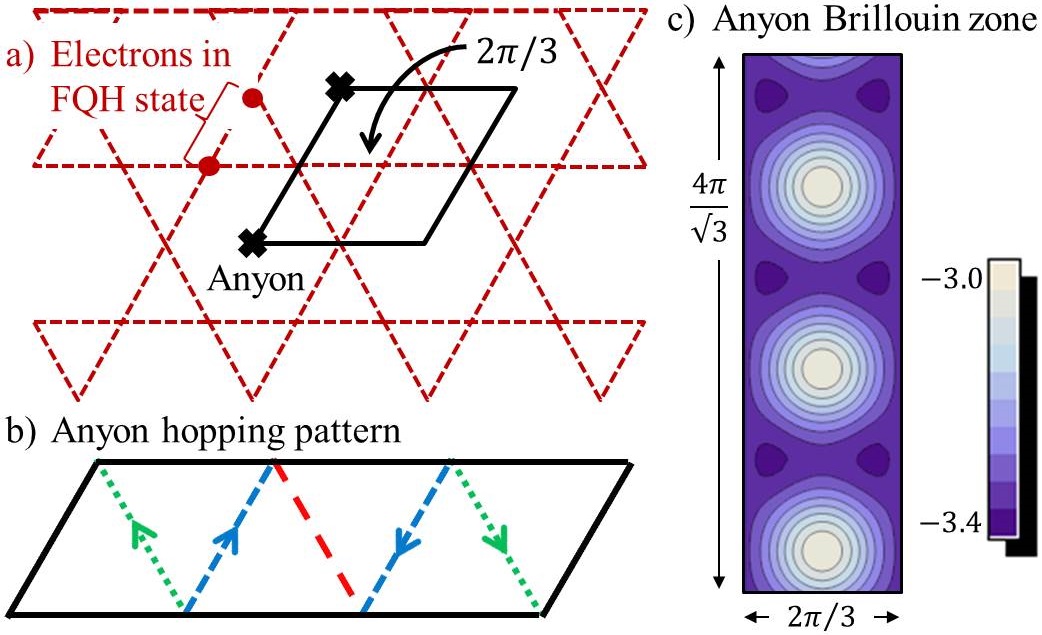}
\caption{(Color online). (a) Anyons live in the center of the hexagons of the
kagome lattice (dashed red lines) to minimize repulsion with the electrons, and
see a flux of $2\pi/3$ per unit cell (solid black line). (b) The anyons hop on
an effective triangular lattice (lattice spacing set to unit width). The
$2\pi/3$ flux breaks translation symmetry by three; here we illustrate a
hopping configuration with uniform flux. In addition to regular $t=-1$ hopping
(solid black lines), a phase of $\pi/3$ is added to the dotted (green) lines in
the direction of the arrow, while $2\pi/3$ is added to the short dashed (blue)
lines with $\pi$ added on the wide dashed (red) line. (c) Resulting band
structure of the lowest band with six degenerate minima.
}\label{fig:anyonlat}
\end{center}
\end{figure}

On a lattice system, anyons have a magnetic length scale on the order of the
lattice spacing \cite{TMW1106} and form a strong local charge distortion,
resulting in an anyon hopping governed by the typical electron hopping energy.
We provide a more detailed discussion and comparison of energy scales in
Appendix \ref{app:scales}, where the relevant anyon energy scales are estimated
as $\sim \hbar^2/m_a l_a^2$ for the kinetic and $\sim(e/3)^2/\epsilon l_a$ for
the interaction energy ($m_a$ and $l_a$ are an effective anyon mass and
interparticle spacing, $\epsilon$ gives the effective screening, e.g. the
dielectric constant of the substrate). Note that the anyon kinetic energy is an
energy scale distinct from the bandwidth of the electron flat band, as the
latter is a delicate balance of several different hopping parameters on a
frustrated lattice; furthermore anyons reside on a separate unfrustrated
lattice (e.g. Fig. \ref{fig:anyonlat}(a)).
When the anyon kinetic energy dominates over the anyon interaction energy, we
will obtain an anyon liquid (while in the opposite limit we expect the anyons
to form a Wigner crystal).

To understand the properties of this anyon liquid, we use a flux attachment
scheme first in a mean-field approximation then with fluctuations.  Several
mean-field states are studied, and some of them are anyon superfluids. As the
anyons are charged, these would be superconductors.
Anyon condensation was first suggested by Laughlin\cite{L8825,L8857} in 1988
and explored by several authors \cite{FHL8979,CWWH8925}. Here we present a
system which provides the novel possibility of the right energy regime to
support such physics.

Our formalism allows several choices of mean-field states, so here we present
three different scenarios.  In the first two, we find that the superfluid
contains quasiparticle excitations with fractionalized statistics, in one case
with a fusion structure of $(\Z_2)^4$ and the other with $(\Z_8)^2$.
The  fractional  statistics of the quasiparticles implies
that the first two superfluids have non-trivial topological order.

Besides these examples of intrinsic topological order, we also present a third
one with purely local excitations, very similar to a BCS-type superconducting
state. The properties and implications of these scenarios are intriguing, and
provide a mechanism for superconductivity driven purely by strong repulsion and
complex hopping of electrons, which opens a new route to
potential high temperature superconductivity.

\section{Host system} We illustrate this on a kagome lattice system with
spin-orbit coupling and time-reversal symmetry breaking \cite{TMW1106}.  Here
we assume that the time-reversal symmetry breaking completely polarizes the
electron spin.\cite{TMW1106} Also, the complex hopping of the electron is such
that as the electron hops around the unit cell anti-clockwise, it gains a phase
$2\pi$ ({\it i.e.} the effective magnetic field $B$ satisfies $-eB>0$ where
$-e<0$ is the electron charge).  A $\nu=1/3$ FQH state in such a system is
roughly described by the Laughlin wave function
\begin{align}
\Psi=\prod_{i<j} (z_i-z_j)^3 \text{e}^{-\frac{1}{4l_B^2}\sum_i |z_i|^2}
\end{align}
where the complex coordinates $z_i$ of the electron live on the kagome lattice.
The state has anyon excitations, and a finite density of anyons is expected to
be created with a small amount of doping (small relative to total electron
density). In order to minimize the energy of the electrons that live on the
kagome lattice, these anyons would reside in the center of the hexagons, see
Fig. \ref{fig:anyonlat}(a).  In the following, we will consider anyons with
charge $-e/3$ ({\it i.e.} quasiparticles instead of quasiholes).

The anyons
pick up a phase of $2\pi$ when moved all the way around an electron
anti-clockwise (as can be seen from the Laughlin wavefunction). Since the
electron density is 1/3 per unit cell, this contributes a $2\pi/3$ phase when
moving around a unit cell anti-clockwise. The anyon lattice translation
vectors thus triple.
In Fig. \ref{fig:anyonlat}(b) we illustrate an anyon hopping with such a flux
configuration and the corresponding bandstructure (Fig. \ref{fig:anyonlat}(c)):
we see six minima which can be labelled with the index $I=1,2,..,6$.
This has a dramatic consequence for the anyon gas: as
these degenerate minima have distinct momentum quantum numbers, our anyons are
now of six species.
The anyons also have
a statistical angle of
$\theta=\pi/3$, {\it i.e.} as we move one anyon half-way around
another anyon anti-clockwise (which corresponds to an exchange),
it will induce a phase $\pi/3$.

We ask, what is the  groundstate
for such a six-species anyon gas?
To understand the ground state, we can describe the anyon with a flux-attachment
procedure \cite{GM8752,ZHK8982,R8986,J8999} where composite fermions $\psi_I$ (also
six species) are attached to a statistical field $a_\mu$ such that the
resulting particles have the appropriate $\theta=\pi/3$ statistics. (Here only
flux from the anyons are included, as the flux giving our underlying $\nu=1/3$
state had been considered earlier --- where it resulted in multiple degenerate
minima for the anyon dispersion hence creating multiple anyon species.) The
Lagrangian is
\begin{eqnarray}
 \mathcal{L}&=&
\frac{\tilde{\nu}}{4\pi} a_\mu\partial_\nu a_\lambda\epsilon^{\mu\nu\lambda}
+i\psi^\dag_I(\partial_0+ia_0-i\frac{e}{3}A_0)\psi_I
\nonumber\\
&+&\frac{1}{2m}|(\partial_i+ia_i-i\frac{e}{3}A_i)\psi_I|^2+...
\label{eq:fermions}
\end{eqnarray}
where $A_\mu$ is the external electromagnetic field, $\tilde{\nu}$
a constant and $m$ the anyon mass. The ``..." denotes other terms that do not
affect the discussion, e.g. the Maxwell term or the Coulomb repulsion between
fermions.

Determining $\tilde{\nu}$ is easier within a hydrodynamic approach
\cite{BW9033,BW9045}, where the low-energy collective modes can be described by
a particle current $j^\mu$
\begin{eqnarray}
 \mathcal{L}&=&
\frac{\tilde{\nu}}{4\pi} a_\mu\partial_\nu a_\lambda\epsilon^{\mu\nu\lambda}
-(a_\mu-\frac{e}{3}A_\mu) j^\mu +\cdots;
\label{eq:hydro}\\
&&j^\mu=\sum_I j^\mu_I,\quad\textrm{and}\quad j^\mu_I=
\frac{1}{2\pi}\partial_\nu \tilde{a}_{\lambda I} \epsilon^{\mu\nu\lambda}
\label{eq:currents}
\end{eqnarray}
since each fermion number current can be associated with a $U(1)$  gauge field.
Introducing a particle that carries an $a_\mu$ unit charge gives the source
term $a_0\delta(\vect{x}-\vect{x_0})$. Varying with respect to $a_0$, we find
this term creates an excitation of charge $Q=-e/\tilde{\nu}$ and is associated
with $1/\tilde{\nu}$ units of the $a_\mu$ flux \cite{Wen04}. Hence,
interchanging two such excitations induces a phase
\begin{eqnarray}
 \pi \times \textrm{(number of }a_\mu\textrm{-flux quanta) }\times (a_\mu \textrm{ charge})=\frac{\pi}{\tilde{\nu}}\nonumber
\end{eqnarray}

There is also a phase $\pi$ from the core statistics of the composite fermions, so these two contributions give the full statistical angle, i.e.
\begin{eqnarray}
\frac{\pi}{\tilde{\nu}}-\pi=\theta\label{eq:tnu}
\end{eqnarray}
Since  $\theta=\pi/3$, we obtain  $\tilde{\nu}=3/4$.

There is simple way to understand the above result: We view the anyon as a
bound state of a fermion and flux $2\pi \frac{4}{3}$. As we move such a bound
state halfway around another bound state anti-clockwise, it induces the
correct statistical phase $\pi \frac{4}{3} -\pi=\pi/3$ where $-\pi$ comes from
the core statistics of the fermions.

\section{Mean-field treatment}
Within this flux-attachment scheme, we use
a mean-field approximation where the statistical flux bound to the composite
fermions is smeared to form a constant background field:
$a_\mu=\bar{a}_\mu+\delta a_\mu$ where the flux density
$\epsilon^{ij}\partial_i\bar{a}_j$ takes a constant average value $b$,  and
$\delta a_\mu=0$.

In this approximation, our anyon gas problem becomes that of fermions in a constant magnetic field $b$. Their resulting ground state depends simply on their filling fraction, which we can calculate from Eq. \ref{eq:fermions} by varying $a_0$:
\begin{eqnarray}
\sum_I\psi^\dag_I\psi_I=\frac{\tilde{\nu}}{2\pi}\epsilon^{ij}\partial_i\bar{a}_j=\frac{\tilde{\nu}b}{2\pi}
\end{eqnarray}
The filling fraction as ratio of electron density $\sum_I\psi^\dag_I\psi_I$ to magnetic field density $b$, is
\begin{eqnarray}
2\pi\frac{\sum_I\psi^\dag_I\psi_I}{b}=\tilde{\nu}=3/4
\end{eqnarray}
The constant $\tilde{\nu}$ in front of our Chern-Simons term has become the
filling fraction of the composite fermions (as distinct from the filling
fraction of our electron system $\nu=1/3$).  In other words, smearing the
$2\pi \frac{4}{3}$-flux per fermion into a constant ``magnetic'' field
induces a positive ``magnetic'' field where the fermions have an effective
filling fraction $\tilde{\nu}=3/4$.

What is the groundstate of a system with six fermion species at a combined filling fraction of 3/4? This would favor a multi-layer analog of the Laughlin state \cite{H8375}:
\begin{eqnarray}
\Psi(\{z_i\})=\prod_{I<J,i,j} (z_i^I-z_j^J)\prod_{I,i<j}^{I=6}(z_i^I-z_j^I)^2e^{-\sum|z_i|^2/4l_B^2}\nonumber
\end{eqnarray}
where $z_i^I$ is the coordinate of the $i$th electron in the $I$th layer, and can be described by a $6\times 6$ $K$-matrix with 3's along the diagonal and 1's on the off-diagonal entries.

Replacing the $\cdots$ term in  Eq.  \ref{eq:hydro} with this term in our
theory (which we denote with $\tilde{K}$ for this composite fermion $K$-matrix)
and substituting Eq. \ref{eq:currents} in Eq.  \ref{eq:hydro}, we obtain
\begin{eqnarray}
 \mathcal{L}&=&
\frac{\tilde{\nu}}{4\pi} a_\mu\partial_\nu a_\lambda\epsilon^{\mu\nu\lambda}
-(a_\mu-\frac{e}{3}A_\mu) \sum_I \frac{1}{2\pi}\partial_\nu
\tilde{a}_{\lambda I} \epsilon^{\mu\nu\lambda}
\nonumber\\
&+&\frac{\tilde{K}_{IJ}}{4\pi} \tilde{a}_{\mu I}\partial_\nu
\tilde{a}_{\lambda J}\epsilon^{\mu\nu\lambda}
\label{eq:fullL}
\end{eqnarray}

\section{Allowing gauge-field fluctuations}
To this mean-field solution, we can now add fluctuations of the gauge-field, i.e. $\delta a_\mu\neq0$. Further, as  $\delta a_\mu=\sum_I\delta \tilde{a}_{\mu I}/\tilde{\nu}$ (as can be seen from varying $a_\mu$), we can substitute this $\delta a_\mu$ gauge-field out. With these steps, the following additional terms due to fluctuations are obtained
\begin{eqnarray}
\delta \mathcal{L} =
(\tilde{K}_{IJ}-\frac{1}{\tilde{\nu}}C_{IJ})\frac{1}{4\pi} \delta\tilde{a}_{\mu I}\partial_\nu\delta \tilde{a}_{\lambda J}\epsilon^{\mu\nu\lambda}
\end{eqnarray}
where $C_{IJ}=1$.
We can introduce an effective $6\times6$ $K$-matrix:
\begin{eqnarray}
\tilde K_{eff}&=&\tilde{K}-\frac{1}{\tilde{\nu}}C
\nonumber\\
&=&
 \begin{pmatrix}
  3 & 1 & \cdots & 1 \\
  1 & 3 & \cdots & 1 \\
  \vdots  & \vdots  & \ddots & \vdots  \\
  1 & 1 & \cdots & 3
 \end{pmatrix}
 - \frac{4}{3}\begin{pmatrix}
  1 & 1 & \cdots & 1 \\
  1 & 1 & \cdots & 1 \\
  \vdots  & \vdots  & \ddots & \vdots  \\
  1 & 1 & \cdots & 1
 \end{pmatrix} .
\label{eq:k}
\end{eqnarray}

It can be verified that the determinant of $\tilde K_{eff}$ above is 0, signifying a zero-mode where the gauge field has no Chern-Simons term and is gapless. To choose a basis where this zero mode is explicit, we employ
\begin{align}
 a'_{\mu I}=(U^{-1})_{IJ}\tilde a_{\mu J},
\end{align}
in which basis the $K$-matrix becomes
\begin{align}
K'=U^T \tilde K_{eff}  U
=\frac{1}{3} \begin{pmatrix}
  5&  -1&  -1&  -1&  -1&  0\\
 -1&   5&  -1&  -1&  -1&  0\\
 -1&  -1&   5&  -1&  -1&  0\\
 -1&  -1&  -1&   5&  -1&  0\\
 -1&  -1&  -1&  -1&   5&  0\\
  0&   0&   0&   0&   0&  0\\
\end{pmatrix}\label{eq:k'}
\end{align}
using
\begin{align}
 U=
\begin{pmatrix}
  0 &  0 &  0&   0&  1&  1\\
 -1 &  0 &  0&   0&  1&  1\\
  0 & -1 &  0&   0&  1&  1\\
  0 &  0 & -1&   0&  1&  1\\
  0 &  0 &  0&  -1&  1&  1\\
  0 &  0 &  0&   0&  0&  1\\
\end{pmatrix}
\in SL(6,\Z)
\end{align}
i.e. we use an invertible integer matrix for $U$ to preserve the integer
quantization of the gauge charges.

In this $a'_{\mu I}$ basis, Eq. (\ref{eq:fullL}) becomes
\begin{eqnarray}
 \mathcal{L}&=&
\frac{K'_{IJ}}{4\pi} a'_{\mu I}\partial_\nu
a'_{\lambda J}\epsilon^{\mu\nu\lambda}
+
\frac{1}{2\pi}
eA_\mu q'_I
\partial_\nu
a'_{\lambda I} \epsilon^{\mu\nu\lambda}
\label{eq:fullLp}
\end{eqnarray}
where
\begin{align}
 q'^T=(-\frac13,-\frac13,-\frac13,-\frac13,\frac53,2)
\end{align}

From Eq. \ref{eq:k'}, we see explicitly that $a'_{\mu 1},\cdots,a'_{\mu 5}$ have Chern-Simons terms and are
gapped, while $a'_{\mu 6}$ does not have a Chern-Simons term and is
gapless. $a'_{\mu 6}$ also
couples to the electromagnetic field $A_\mu$ (since $q'_6\neq0$), so $a'_{\mu
6}$ is the only gapless mode describing charge density fluctuations.

\begin{figure}[tb]
\begin{center}
\includegraphics[width=\linewidth]{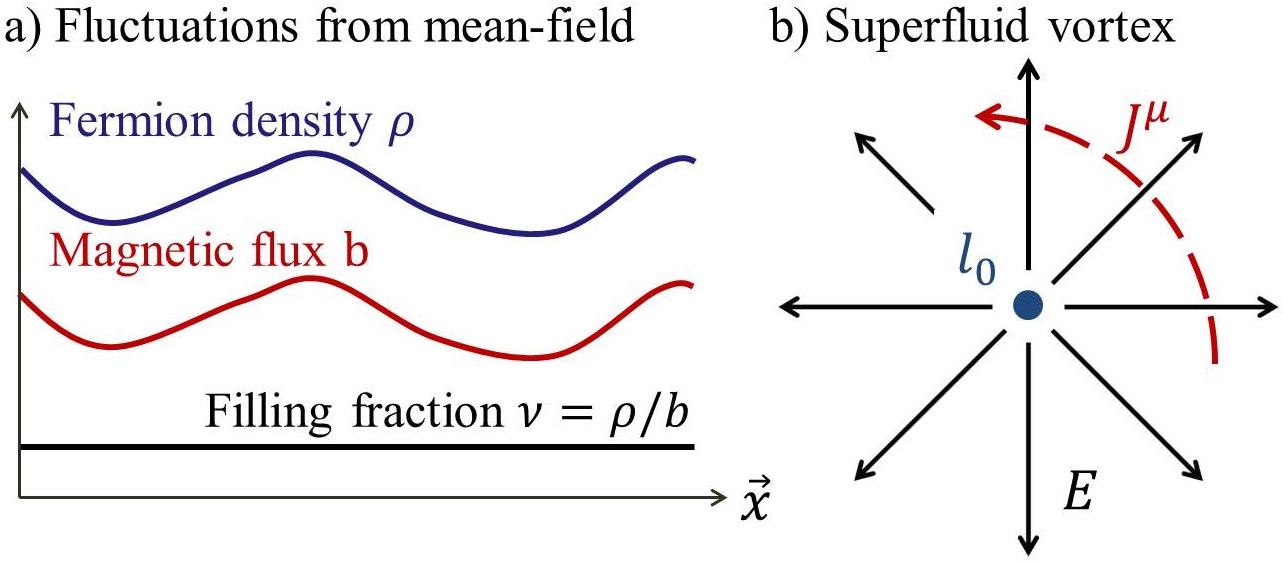}
\caption{(Color online). (a) Heuristic scheme of the superfluid gapless mode. Given a mean-field quantum Hall state, as the fermion density fluctuates, so does the flux density since the
flux-attachment construction combines fermions and flux in a particular
ratio. Hence density fluctuations do not alter the ratio of fermion to flux
density, leaving the filling fraction constant everywhere. This leaves the
mean-field FQH groundstate of the composite fermions locally intact everywhere,
so the density mode is a gapless excitation while the rest of the system
remains gapped -- forming a superfluid. (b) Vortex quantization of the
superfluid current $J_{\mu}$, created by the electric field $E$ emanating from
the charge $l_0$ in the dual picture.
}
\label{fig:stuff}
\end{center}
\end{figure}

The presence of a gapless mode can be understood heuristically from Fig.
\ref{fig:stuff} (a). Given a mean-field quantum Hall state, any fluctuations of the fermion density and flux density always co-fluctuate in
the same way (in this flux-attachment construction). Hence their ratio, the filling fraction, remains constant locally.
As the mean-field quantum Hall state remains the true local solution
everywhere, any density fluctuations have vanishing energy, forming a gapless mode. With one gapless density mode and all other
excitations gapped, this implies the state is a charged
superfluid/superconductor.\cite{WZ9040,WZ9211}


\section{Superfluid properties}
We examine two aspects of the superfluid: the
nature of the gapless mode, and gapped quasiparticle excitations.

For the former, this collective mode has a definite vortex quantization and can
be calculated using 2+1D duality between the XY-model and $U(1)$ gauge theory,
since the unit gauge charge of $a'_{\mu 6}$ corresponds to the superfluid vortex
(see Fig. \ref{fig:stuff} (b) and Appendix \ref{app:vortex} for details). We
find that $q'_6=2$ implies the flux quantization is $\pi/e$, corresponding to $hc/2e$ in cgs units -- reminiscent of BCS-like electron pair
condensation with an order parameter $<c c>\neq 0$.  However, even if we
identify an order parameter $<c c>\neq 0$, we
cannot conclude that the anyon superconductor discussed above is a BCS
superconductor.  In fact, the following discussion reveals clearly non-BCS-like
properties of the above  anyon superconductor.

We proceed to analyze properties of the quasiparticles with \emph{finite}
energy gaps.  Those excitations are described by integer charges of the
$a'_{\mu 1},\cdots,a'_{\mu 5}$ gauge fields.\cite{Wen04}  The  finite
Chern-Simons terms for these  gauge fields lead to short-ranged interactions
and finite energy of their gauge charges. Isolating the $5\times5$ gapped subspace of $K'$ from Eq. \ref{eq:k'},
\begin{align}
  K'_{5\times 5}
=\frac{1}{3} \begin{pmatrix}
  5&  -1&  -1&  -1&  -1\\
 -1&   5&  -1&  -1&  -1\\
 -1&  -1&   5&  -1&  -1\\
 -1&  -1&  -1&   5&  -1\\
 -1&  -1&  -1&  -1&   5\\
\end{pmatrix},
\end{align}
we use five-dimensional
$l'$-vectors to describe these integer charges of the $a'_{\mu
1},\cdots,a'_{\mu 5}$ gauge fields, as all finite energy excitations can be
labeled by $l'$-vectors.

In this case, the mutual statistical angle\cite{Wen04} between any $l'_1$
and $l'_2$ is given by
\begin{align}
2\pi \bm{l'_1}^T K_{5\times 5}^{' -1} \bm{l'_2}.
\end{align}
The statistical angle and electric charge of a single $\bm{l'}$ excitation are given respectively by
\begin{align}
 \theta_{\bm{l'}}=\pi \bm{l'}^T K_{5\times 5}^{'-1} \bm{l'},\ \ \
 Q_{\bm{l'}} = \bm{q'}^T K_{5\times 5}^{'-1} \bm{l'}.
\end{align}
We note that this electric charge of an $\bm{l'}$ is only defined
up to an even integer due to the $2e$ charge condensation.


A subset of  $l'$-vectors describes  excitations with trivial mutual
statistics among themselves, and are bosons with even electric charge or
fermions with odd electric charge.  They can be viewed as
bound states of electrons and are topologically trivial excitations. To obtain a complete basis for this subset, we decompose
\begin{align}
  K_{5\times 5}^{'-1} &=\frac12
 \begin{pmatrix}
  2&   1&   1&   1&   1\\
  1&   2&   1&   1&   1\\
  1&   1&   2&   1&   1\\
  1&   1&   1&   2&   1\\
  1&   1&   1&   1&   2\\
\end{pmatrix}
\nonumber\\
&=U \Lambda V^{-1}
\end{align}
with $U,V\in SL(5,\Z)$
\begin{align}
U &=
\begin{pmatrix}
 -1 &-7& -6 &-10 &-11\\
  0  &0 & 0  &-1  & 0\\
  0  &0 &-1  &-4  & 0\\
  0  &1 & 1  &-1  & 0\\
  0  &0 & 0  &-2  &-1\\
\end{pmatrix} ,
\nonumber\\
\Lambda &=
\begin{pmatrix}
 3 &0 &0 &0 &0\\
 0 &\frac12 &0 &0 &0\\
 0 &0 &\frac12 &0 &0\\
 0 &0 &0 &\frac12 &0\\
 0 &0 &0 &0 &\frac12\\
\end{pmatrix},
\nonumber\\
V&=
\begin{pmatrix}
 -5 &-6 &-5 &-7 &-9\\
  1 & 1 & 1 & 2 & 2\\
  1 & 1 & 0 &-1 & 2\\
  1 & 2 & 2 & 2 & 2\\
  1 & 1 & 1 & 1 & 1\\
\end{pmatrix}.
\end{align}
We find that the set of $\bm{l'}$-vectors describing trivial excitations is generated by the first column of $V$ and two times the second to
the fifth columns of $V$:
\begin{align}
 \bm{l'}^T_{\textrm{triv}} &=
(-5,1,1,1,1),
(-12,2,2,4,2),
(-10,2,0,4,2),
\nonumber\\
&\ \ \ \
(-14,4,-2,4,2),
(-18,4,4,4,2).
\end{align}
The above basis vectors can be simplified to
\begin{align}
 \bm{l'}^T_{\textrm{triv}} &=
(-5,1,1,1,1),
(2,-2,0,0,0),
(2,0,-2,0,0),
\nonumber\\
&\ \ \ \
(2,0,0,-2,0),
(2,0,0,0,-2).
\end{align}
Since $(q^{'T} U\Lambda )^T = (1,1,1,1,1)$, the first vector describes a fermion (odd charge), while all others are bosons (even charge).

The $\bm{l'}$-vectors not in this subset describe topological excitations.
Two $\bm{l'}$-vectors differing by an $\bm{l'}$-vector in the trivial subset are regarded
as the same type of topological excitation.  We find there are 16 types
of  topological excitations (including the trivial type). They are described by
\begin{eqnarray}
l'_{\alpha\beta\gamma\delta}&=&
\alpha \bm{l'}_{1000}+
\beta \bm{l'}_{0100}+
\gamma \bm{l'}_{0010}+
\delta \bm{l'}_{0001},
\nonumber\\
\bm{l'}_{1000}&=&(1,-1,0,0,0)^T,\ \ \ \ \
Q_{1000}=0,\nonumber\\
\bm{l'}_{0100}&=&(1,0,-1,0,0)^T,\ \ \ \ \
Q_{0100}=0,\nonumber\\
\bm{l'}_{0010}&=&(1,0,0,-1,0)^T,\ \ \ \ \
Q_{0010}=0,\nonumber\\
\bm{l'}_{0001}&=&(1,0,0,0,-1)^T, \ \ \ \ \
Q_{0001}=1.
\label{eq:fourl}
\end{eqnarray}
where $\alpha,\beta,\gamma,\delta=\{0,1\}$ only.
%
%
%
Composites of two identical particles are trivial, so each of the four indices
take only 0 or 1 to be actually different types of
excitations (there is a $\Z_2$ fusion
structure for each, separately). However, a composite of two different
particles, e.g. $\bm{l'}_{1100}=\bm{l'}_{1000}+\bm{l'}_{0100}$ is another
non-trivial excitation distinct from the underlying two, giving rise to
$2^4=16$ possible combinations in total -- $\bm{l'}_{0000}$ is trivial and the
other 15 are not. Hence the fusion relations between these particles have a
group structure of $(\Z_2)^4$.


\subsection{Intrinsic topological order}
The 16-by-16 modular matrix
$S_{ab}=\frac{1}{\sqrt{D}}\textrm{exp}(\text{i}2\pi
l_{a}^{'T}K^{'-1}_{5\times 5}l'_{b})$ ($D$ is the quantum dimension for
normalization \cite{Wang10}) for all 16 quasiparticle types can be calculated,
along with the diagonal twist matrix $T_{ab}=\delta_{ab}\textrm{exp}(\text{i} \pi l_{a}^{'T}K^{'-1}_{5\times 5}l'_{b})$.  We illustrate $S$ and $T$ for just the four
generating vectors listed in Eq. \ref{eq:fourl}:
\begin{eqnarray}
S&=& \frac{1}{4}\begin{pmatrix}
  1 & -1 & -1 & -1 &\cdots\\
  -1 & 1 & -1 & -1 &\cdots\\
  -1  & -1  & 1 & -1  &\cdots\\
  -1 & -1 & -1 & 1 &\cdots\\
\vdots &\vdots &\vdots &\vdots &\ddots \\
 \end{pmatrix},\nonumber\\
 T&=& \begin{pmatrix}
  -1 & 0 & 0 & 0 &\cdots\\
  0 & -1 & 0 & 0 &\cdots\\
  0  & 0  & -1 & 0  &\cdots\\
  0 & 0 & 0 & -1 &\cdots\\
\vdots &\vdots &\vdots &\vdots &\ddots \\
 \end{pmatrix};
\end{eqnarray}
where the -1 entries in $S$ represent mutual semion statistics. The -1 entries in $T$ denote fermion statistics, an example of spin-charge separation since we noted in Eq. \ref{eq:fourl} that three of these four excitations carry zero charge.

We find that these matrices satisfy the modular group relations as expected for a bosonic topological order, e.g.
\begin{eqnarray}
(ST)^3=\exp{\frac{\pi i c}{4}}S^2\label{eq:st3}
\end{eqnarray}
where the ``statistical'' central charge $c$ is 4.

This superfluid phase above of the six-species anyon gas has six branches of edge
modes (central charge $c=6$) which all move in the same direction.  Five
branches of edge modes come from the $a'_{\mu I},\ I=1,\cdots,5$ gauge fields
that have non-zero Chern-Simons terms (the other gauge field $a'_{\mu 6}$
corresponds to the gapless bulk density mode).  The sixth edge mode
comes from the underlying $\nu=1/3$ FQH state.

\subsection{Other possible scenarios}
Our method depends upon particular choices
of parameters and this previous example is just the simplest choice. Here we examine
other possible outcomes within this scheme. For instance, we could view the anyon
as a bound state of a fermion and flux $-2\pi \frac{2}{3}$. As we move such a
bound state halfway around another bound state anti-clockwise, it will also
induce the correct statistical phase $-\pi \frac{2}{3} +\pi=\pi/3$ where $+\pi$
comes from the core statistics of the fermions.
If we smear the
$-2\pi \frac{2}{3}$-flux per fermion into a constant ``magnetic'' field,
the  ``magnetic'' field will
be negative, and the fermions will have an effective
filling fraction $\tilde{\nu}=3/2$.
In this case, our six composite fermions would have a combined filling fraction of 3/2, where a favourable groundstate could be three Halperin states each at filling fraction 1/2.

The Halperin wavefunction is expected to be the groundstate of a bilayer system at filling fraction 1/2 \cite{H8375}:
\begin{eqnarray}
\Psi_H(\{z_i\})=\prod_{I<J,i,j} (z_i^{*I}-z_j^{*J})\prod_{I,i<j}^{I=2}(z_i^{*I}-z_j^{*I})^3e^{-\sum|z_i|^2/4l_B^2}\nonumber
\end{eqnarray}
when the intra-layer repulsion is stronger than the inter-layer repulsion.
The wave function depends on $z_i^{*I}$ since the effective ``magnetic'' field is negative.
The $K$-matrix is three copies of the $K_H$-matrix for a bilayer system:
\[K_H= \left( \begin{array}{cc}
3 & 1  \\
1 & 3  \end{array} \right),\quad\tilde{K}=\left( \begin{array}{ccc}
K_H & 0&0  \\
0&K_H&0\\
0 & 0&K_H  \end{array} \right)\]

Repeating a similar analysis, we find once again a zero-mode with the same superfluid vortex quantization. The possible gapped quasiparticles are now generated by just two $\{l_{\alpha\beta}\}$-vectors
\begin{eqnarray}
l_{\alpha\beta}&=& \alpha \bm{l}_{10}+ \beta \bm{l}_{01},
\nonumber\\
\bm{l}_{10}&=&(0,1,-1,0,0,0)^T,\nonumber\\
\bm{l}_{01}&=&(-1,0,2,0,0,-1)^T.\label{eq:l2}
\end{eqnarray}
but now $\alpha,\beta=\{0,1,...,7\}$, i.e. each excitation has a separate
fusion structure of $\Z_8$. Together, they form $8^2=64$ possible combinations
and have a combined fusion structure of $(\Z_8)^2$.

As above, we can compute their mutual statistics using the appropriate $K^{-1}_{\textrm{eff}}$, which here has $1/3$ along the diagonal, $-1/6$ on off-diagonal entries within each bilayer and $-1/24$ on all remaining off-diagonal entries. Again we present the $S$ and $T$ matrices for the two generating vectors in Eq. \ref{eq:l2}:
\begin{eqnarray}
S= \frac{1}{8}\begin{pmatrix}
  -i & \exp{\frac{3i\pi}{4}}  &\cdots \\
  \exp{\frac{3i\pi}{4}} & i &\cdots\\
 \vdots &\vdots &\ddots \\
 \end{pmatrix},T= \begin{pmatrix}
  \exp{\frac{3i\pi}{4}} & 0 &\cdots \\
  0 & \exp{\frac{i\pi}{4}} &\cdots\\
 \vdots &\vdots &\ddots \\
 \end{pmatrix}.\nonumber
\end{eqnarray}

Besides the strangeness of the statistics obtained, we find the relation in Eq.
\ref{eq:st3} is violated as $(ST)^3=-\mathbb{I}$ whereas $S^2$ is an
off-diagonal matrix (while $S^4=\mathbb{I}$). Hence this is not a bosonic
topological order, and it is unclear if this topological order that comes with
a gapless mode has the same properties as topological order of a fully gapped
system.

This second superfluid phase of the six-species anyon gas has six branches of
edge modes, five of them moving in one direction and the other moving in the
opposite direction.  The five branches of edge modes moving in the same
direction come from the five $\tilde a_{\mu I}$ gauge fields that have non-zero
Chern-Simons terms (the other combination of $\tilde a_{\mu I}$ gauge fields
corresponds to the gapless bulk density mode).  The sixth edge mode
moving in the opposite direction comes from the underlying $\nu=1/3$ FQH state.
It is possible that the interaction between edge modes may reduce
them into four branches moving in same direction.

\subsection{BCS-like state}
The last scenario we present is a case with particularly simple results, with properties similar to that obtained from BCS theory. If the anyon hopping is frustrated, this adds a minus sign to all the anyon hoppings in Fig. \ref{fig:anyonlat}(b) and the lowest band becomes the flipped version of Fig. \ref{fig:anyonlat}(c). What were previously three maxima become the location of three minima, which results in just three composite fermion species instead of six.

Considering the filling fraction $\tilde{\nu}=3/2$, we see that if all three fermions have the same density, they each have a filling fraction of
 1/2, which is a compressible state.

However, if the lattice translation symmetry is broken by spontaneous formation of a charge-density wave or by the application of a periodic electrostatic potential, this could suppress some of the fermion species density relative to others. Here we choose to work in an alternate Wannier basis where the index $I$ for fermion species now denotes fermion species in real and not momentum space \cite{Q1103}.

\begin{figure}[tb]
\begin{center}
\includegraphics[width=\linewidth]{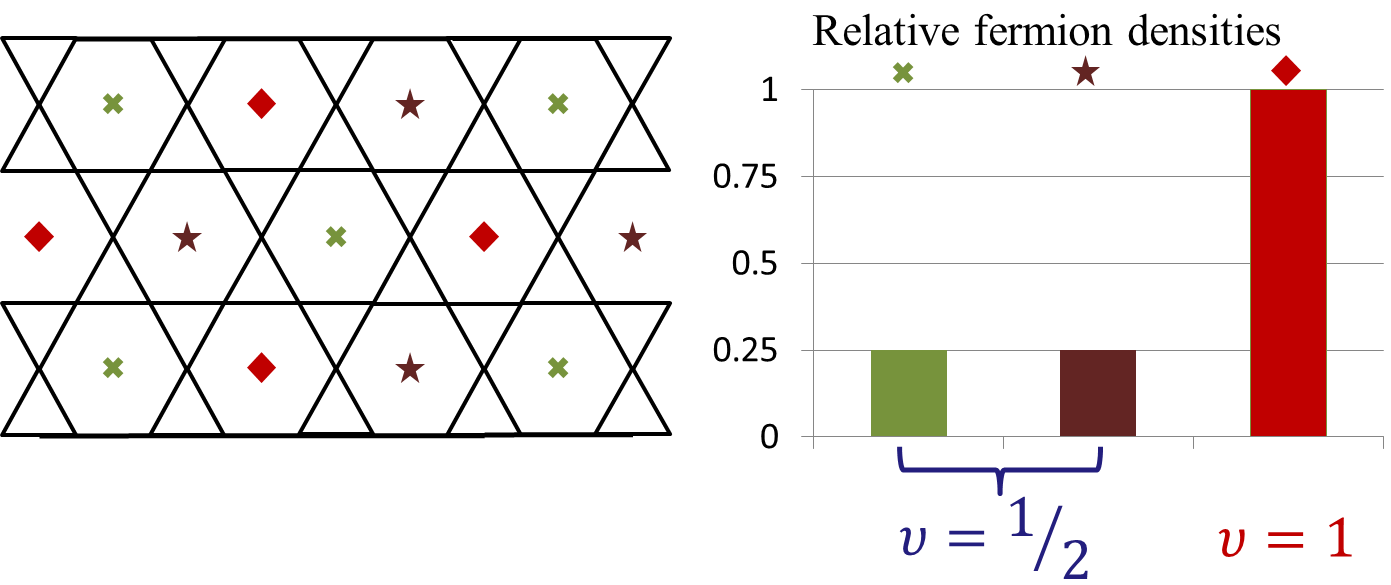}
\caption{(Color online). (a)Breaking of translation symmetry in the lattice, where the three species (cross, star and diamond) have different relative densities. This could happen by spontaneous formation of a charge-density wave or by the application of a periodic electrostatic potential. (b) When the first two species have a relative density of 1/4 compared to the third, this favors a Halperin state for the first two species and an integer quantum Hall state for the third.}\label{fig:3anyons}
\end{center}
\end{figure}

 With a charge imbalance where two species have a relative density of 1/4 compared to the third (see Fig. \ref{fig:3anyons}), this would be a Halperin state for the first two species and an integer quantum Hall state for the last.
This is described by the $K$-matrix
\[\tilde{K}= \left( \begin{array}{cc}
K_H & 0 \\
0  & 1  \end{array} \right)\]
and has a zero mode with the same $2e$ quantization like in previous examples, with quasiparticle excitations that are non-fractionalized.

These purely local quasiparticle excitations can be described with the $\bm{l}$-vectors and topologically trivial $S$ and $T$ matrices
\begin{eqnarray}
\bm{l}_1&=&(-2,-2,1)^T,\quad \bm{l}_2=(-1,1,0)^T; \nonumber\\
S&=& \frac{1}{\sqrt{2}}\begin{pmatrix}
  1 & 1 \\
  1 & 1
 \end{pmatrix},\quad T=\begin{pmatrix}
  -1 & 0 \\
 0 & -1
 \end{pmatrix}.
\end{eqnarray}

This third superfluid phase for the three-species anyon gas has three branches of
edge modes, two of them move in one direction and the other moves in the
opposite direction.  The two branches of edge modes in the same direction
come from the two $\tilde a_{\mu I}$ gauge fields with non-zero
Chern-Simons terms.  The third edge mode moving in the opposite
direction comes from the underlying $\nu=1/3$ FQH state.  It is possible
that the interaction between edge modes may reduce them into one branch
of edge mode, agreeing with the edge mode of the $d+i d$ BCS superconductor.

\section{Discussion}
We show that flat band systems which support a FQH state at commensurate filling, could support a superfluid mode at incommensurate filling. Such a state may have intrinsic topological order, and we present one example of bosonic topological order with anyon fusion statistics of $(\Z_2)^4$ and a second non-bosonic topological order with fusion statistics of $(\Z_8)^2$.
Another possible outcome is a state similar to that from BCS theory, which suggests that such a state could also be described using more direct methods like mean-field theory.

While our model has been based on a kagome lattice, our results essentially rest on the effects of an underlying lattice where the FQH state can be realized. A different route to the same physics is through application of a periodic potential in other continuum-like FQH systems, including semiconductors or graphene.

In order to identify which groundstate has the lowest energy, further work is needed. Besides numerical simulations, the results here suggest anyon wavefunctions and more indirectly, electron wavefunctions, that can be useful in suggesting compatible Hamiltonians or appropriate variational wave-functions. It is also of interest to be able to detect such a superfluid state, and tunnelling into the edge modes may reveal its edge properties and help with its identification.

We thank Steve Kivelson and Zhenghan Wang for helpful discussions.
This research is supported by NSF Grant No.
DMR-1005541, NSFC 11074140, and NSFC 11274192.  Research at Perimeter Institute
is supported by the Government of Canada through Industry Canada and by the
Province of Ontario through the Ministry of Research.

\appendix

\section{Estimation of anyon energy scales\label{app:scales}}
The anyon interaction energy can estimated by the Coloumb repulsion between them, $\sim(e/3)^2/\epsilon l_a$. Their charge is $e/3$ and $l_a$ is their interparticle spacing as determined by their density. $\epsilon$ gives the effective screening, e.g. is the dielectric constant of the underlying substrate.
\begin{figure}[tb]
\begin{center}
\includegraphics[width=0.7\linewidth]{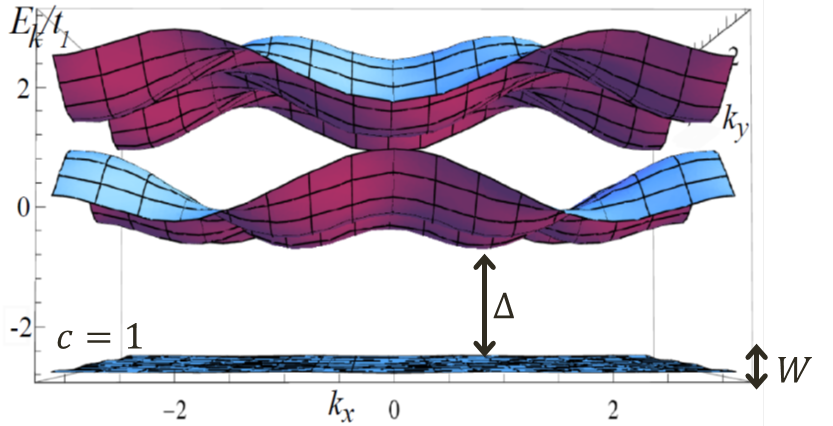}
\caption{(Color online). Illustration of relevant energy scales in the flat band \cite{TMW1106}: $\Delta$ is the bandgap and on the order of the typical electron hopping $t$, which we expect to govern the anyon hopping strength. This is distinct from the width of the flat band $W$ --- a finely tuned balance of several hopping and spin-orbit parameters.}\label{fig:flatband}
\end{center}
\end{figure}

As for the anyon hopping energy, we first look at Fig. \ref{fig:flatband} for a reminder of key energy scales in this system: $\Delta\sim t$ is the typical electron hopping and $W$ is the bandwidth of the flat band. As mentioned in the introduction, since an anyon has a magnetic length scale on the order of the lattice spacing here \cite{TMW1106}, it forms a strong local charge distortion (in contrast to anyons from a FQH state in semiconductor systems with much larger magnetic length scales hence only weakly distorting the wavefunction). In our case, since the anyon/electron interaction energy scale is similar to or larger than the bandgap $\Delta$, the presence of an anyon can cause significant interband mixing. In this case, the anyon hopping will be determined by the typical electron hopping (also the scale of $\Delta$), giving rise to an effective anyon mass $t\sim\hbar^2/m_a a^2$; here $a$ is the lattice spacing.

 Using this effective anyon mass $m_a$, we can now estimate the anyon kinetic energy as  $\sim \hbar^2/m_a l_a^2$. This is an energy scale distinct from $W$, as the latter is a fine balance of different hopping parameters (e.g. the typical electron hopping and spin-orbit coupling) on a frustrated lattice; besides, the anyons reside on a separate unfrustrated lattice. The regime we are interested in is where the anyon kinetic energy dominates the anyon interaction energy.

\section{Vortex quantization\label{app:vortex}}

The gapless density mode in the superfluid is described by
$a'_{\mu 6}$ in Eq. (\ref{eq:fullLp}) and has the Lagrangian
\begin{align}
  \mathcal{L}=la_0\delta(\vect{x})+\frac{1}{2g}e_i^{2}-\frac{1}{2g'}b^{2}\nonumber\\
 +eq' \frac{1}{2\pi}A_\mu\partial_\nu a_{\lambda } \epsilon^{\mu\nu\lambda}+...
\end{align}
where $a_\mu=a'_{\mu 6}$, $q'=2$, $e_i=\partial_0a_i-\partial_i a_0$ and
$b=\partial_1a_2-\partial_2 a_1$.
The $l$-unit of $a_\mu$ charge $la_0\delta(\vect{x})$ corresponds to $l$-unit
of vortex in the superfluid.

What is the vorticity of  $l$-unit of vortex?
%
%
Varying with respect to $a_0$, we obtain Gauss's Law:
\begin{eqnarray}
\nabla\cdot\vect{e}=g l_0\delta(\vect{x})
\end{eqnarray}
This gives an electric field (see Fig. \ref{fig:stuff}(b))
\begin{eqnarray}
\vect{e}=\frac{gl}{2\pi}\frac{\vect{x}}{x^2}\label{eq:efield}
\end{eqnarray}
that creates a density current since
\begin{eqnarray}
J^i=\frac{1}{e}\frac{\partial\mathcal{L}}{\partial A_i}&=&\frac{q'}{2\pi}\partial_\nu a_0 \epsilon^{\mu\nu 0}\nonumber\\
&=&\frac{q'}{2\pi}e_j \epsilon^{ij 0}.
\label{eq:current}
\end{eqnarray}
Combining this with the radially-directed electric field in Eq. \ref{eq:efield}, we obtain
\begin{eqnarray}
\vect{J}=\frac{q'gl}{(2\pi)^2}\frac{\hat{\theta}}{|\vect{x}|}
\end{eqnarray}
i.e. a circulating current around the charge couples to the probe field $A_\mu$ (a vortex as expected from the superfluid/$U(1)$ duality in 2+1 dimensions).

This vorticity is quantized, as we can see by integrating the current around a
loop
\begin{eqnarray}
\oint d\vect{x}\cdot \vect{J}\frac{m}{\rho}&=&\oint d\vect{x}\frac{l}{q'}\frac{\hat{\theta}}{|\vect{x}|}
=2\pi\frac{l}{q'}
\label{eq:vortexquant}
\end{eqnarray}
where $m$ and $\rho$ are the mass and density of superfluid particles.
Their ratio can be converted to a quantity involving the gauge field couplings
$g$ and $q'$ by comparing the dual terms in the action: The kinetic terms in
the action for both the superfluid and $U(1)$ descriptions, $e_i^{2}/2g$ and
$\frac{1}{2}mv^2\rho=\frac12 J^2m/\rho$ ($v$ is the superfluid velocity), can
be converted into each other using $J=v\rho$ and Eq. \ref{eq:current}. This
gives $m/\rho=(\frac{2\pi}{ q'})^2/g$, which we use to obtain the result in Eq.
\ref{eq:vortexquant}.

In all our examples,
the vorticity quantization obtained is $2\pi/q'=\pi$ (since $q'=2$), which corresponds to
a superconducting flux quantization of $hc/2e$, similar to that in a BCS-type
superconductor.

\bibliography{mybib}


\end{document}